\documentclass[a4paper]{article}
\usepackage{times}
\usepackage{graphicx}

\def\la{\;\raise0.3ex\hbox{$<$\kern-0.75em\raise-1.1ex\hbox{$\sim$}}\;}
\def\ga{\;\raise0.3ex\hbox{$>$\kern-0.75em\raise-1.1ex\hbox{$\sim$}}\;}

\begin{document}

\fontsize{14pt}{14pt}\selectfont

%{\setlength{\parindent}{0pt}
\thispagestyle{empty}

\begin{center}

{\Large\bf
HD Molecular Lines \\
in an Absorption System at Redshift ~z = 2.3377 \\
}

\bigskip
\bigskip

{\large  D.A.~Varshalovich$^1$, A.V.~Ivanchik$^1$, \\ P.~Petitjean$^2$,
         R.~Srianand$^3$, C.~Ledoux$^4$}
	 
	 \bigskip
	 \bigskip
	 
$^1$ Ioffe Physical Technical Institute RAS, St.-Petersburg \\

$^2$ Institut d'Astrophysique de Paris -- CNRS, France \\
$^3$ IUCAA, Pune, India \\
$^4$ ESO, Munchen, Germany \\

\end{center}

\bigskip
\bigskip
\bigskip

We have analyzed the spectrum of the quasar PKS~1232+082 obtained by
Petitjean et al. (2000). HD molecular lines are identified in an
absorption system at the redshift z$\,$=$\,$2.3377. 
The column density of HD molecules in the system is estimated,
$N(\mbox{HD})=(1-4)\cdot 10^{14}$~cm$^{-2}$.
The temperature of excitation of the first rotational level
$J=1$ relative to the ground state $J=0$ is $T_{ex}=(70\pm7)$~K.
This is, to our knowledge, the first detection of HD molecules at
high redshift.

\bigskip
\bigskip

{\it Keywords:}~~ quasar spectra, observational cosmology

\vspace{4cm}

Accepted for publication in ``Astronomy Letters''

\newpage

\subsection*{INTRODUCTION}

\hspace{1cm}	 
    The relative abundance of deuterium [D]/[H] formed 
during Big Bang nucleosynthesis is one of the key parameters of 
contemporary cosmology because it is the most sensitive indicator 
of the baryon density in the Universe.

  Deuterium abundance at early cosmological epoch (10-14 Gyrs ago)
may be determined from high-redshift quasar spectra. Until now, 
only atomic lines of D~I  and H~I have been used for measuring 
relative abundances of D~I and H~I. However, this method meets 
serious difficulties. D~I and H~I wavelengths almost coincide, 
viz $\lambda (\mbox{H~I}) / \lambda (\mbox{D~I}) = 1.00027$. 
Moreover the column densities of DI and HI differ by a factor 
of $10^{-4} - 10^{-5}$. 
Therefore, D~I lines are very weak and practically undetectable 
when the column density of hydrogen is low. 
In case the hydrogen column density is high enough, H~I lines are 
saturated and broadened, so that D~I lines are lost in the H~I 
lines. In addition, one cannot be sure that the lines treated as 
D~I lines are actually D~I and not produced by an intervening 
weak H~I cloud shifted in velocity by $-80$~km/s. 
(The latter explanation is possible because the lines in question 
are situated within Ly-$\alpha$ forest where absorption features 
are numerous). The above difficulties may be the reasons 
why [D]/[H] values determined from the atomic lines by different 
authors differ by more than one order of magnitude. 
For example, $2\cdot 10^{-4}$ (Webb et. al, 1997), 
and $1.6\cdot 10^{-5}$ (Pettini \& Bowen, 2001).

    The above difficulties do not arise if one measures
the relative abundances of molecules [HD]/[H$_2$] since the
appropriate wavelengths differ considerably while the redshift 
parameter is the same. It may be difficult however to derive 
[D]/[H] from [HD]/[H$_2$] because of uncertainties in the 
chemistry. Nevertheless this is an independent access to the 
deuterium abundance in these remote clouds.
    Up to now HD lines have not been identified in quasar
spectra. Moreover, for a long period, from 1985 to 1997, the only
molecular absorption system was known, viz H$_2$ system at
$z_{abs} = 2.811$ imprinted in the spectrum of PKS~0528-250
(Levshakov \& Varshalovich, 1985). Today only four reliable 
absorption systems of molecular hydrogen H$_2$ are known
in quasar spectra (Petitjean et al., 2000; Levshakov et al., 2001)
and no other molecules are detected in their optical spectra.

\bigskip

\subsection*{RESULTS OF ANALYSIS OF PKS~1232+082 SPECTRUM}

\hspace{1cm}	 
Here we report on identification of HD lines of absorption
system at $z_{abs} = 2.3377$ in the spectrum of PKS~1232+082
($z_{em} = 2.57$ and $m_V = 18.4$). It is the first
identification of redshifted HD molecules, to our knowledge.  

    The high resolution spectrum of PKS~1232+082 was observed 
using  UVES with the 8.2-m Telescope VLT of ESO 
by Petitjean et al., (2000). The spectrum contains a strong
absorption system of molecular hydrogen H$_2$ at 
$z_{abs}=2.3377$ found earlier by Ge \& Bechtold (1999).

    We have found several HD lines of Lyman series
$B\,^1\Sigma^+ - X\,^1\Sigma^+$ corresponding to R(0) transitions
from the ground state $J=0$, $v=0$, and some tentative lines R(1)
from the first rotational level $J=1$, $v=0$.
    Fig.1 shows fragments of the observed spectrum of PKS~1232+082,
and our synthetical fit of HD lines. One can see
R(0) lines in L~5-0, L~4-0, L~3-0, and L~0-0 bands and some possible
R(1) lines whereas HD lines in L~2-0 and L~1-0 bands are heavily
blended by H$_2$ lines. Parameters of identified HD lines
are presented in Table~1. The spectroscopic data of laboratory lines
are taken from measurements by Dabrowski \& Herzberg (1976) and 
oscillator strengths by Allison \& Dalgarno (1970).

The weighted mean value of the redshift parameter for the HD absorption
system measured reads
\[
z_{abs}(\mbox{HD})=2.337700(5) \, ,
\]
in good agreement with $z_{abs}(\mbox{H$_2$})=2.33771$ 
(Petitjean et al., 2000).

According to our estimates, the column densities of HD molecules 
in the ground state $J=0$ and the first rotational state $J=1$ 
are
\[
N_{J=0}(\mbox{HD})\,=\,(1 - 3)\cdot 10^{14} \; \mbox{cm}^{-2} \, ,
\]
\[
N_{J=1}(\mbox{HD})\,=\,(4 - 8)\cdot 10^{13} \; \mbox{cm}^{-2} \, .
\]

The population of the first rotational level relative to the 
ground state may be characterized by the excitation temperature:
\[
T_{ex}\,=\,70\pm7 \; \mbox{K} \, .
\]

However, the R(1) lines have low S/N ratios, so that this value 
may be considered as an upper limit of $T_{ex}$. New observations
with higher S/N are necessary to confirm the lines from the J=1 level.

A detailed analysis of the relative abundances of [HD]/[H$_2$]
and corresponding estimates of [D]/[H] will be done elsewhere.
Note that an additional problem of interstellar
chemistry (concerned with formation and destruction of the 
molecules) has to be solved for 
the determination of [D]/[H] from [HD]/[H$_2$]. 

In conclusion, we emphasize that the detection of HD molecules
in absorbing matter at such high redshift  may be important to
understand the formation of the first generation of stars 
because HD molecules might be important cooling agents in 
the primordial condensations where heavy elements were in deficit

\bigskip
\bigskip

{\it Acknowledgment:} Work of DAV and IAV was partly supported by
the RFBR (99-02-18232, 01-02-06098) and the Russian State Program 
``Astronomy''. The observations have been obtained with UVES 
mounted on the 8.2~m KUYEN telescope operated by the European Southern
Observatory at Paranal, Chile. RS and PPJ acknowledge support from the
Indo-French Centre for the Promotion of Advanced Research (Centre
Franco-Indian pour la Promotion de la Recherche Avanc\'ee) under
contract number 1710-1.
                     
\newpage

\subsection*{REFERENCES}

Allison~A.C., Dalgarno~A. // Atomic Data, 1970, V. 1, P. 289. \\
Dabrowski~I., Herzberg~G. // Canadian J. Phys., 1976, V. 54, P. 525. \\
Flower D.R., Roueff E. // MNRAS, 1999, V. 309, P. 833. \\ 
Ge~J., Bechtold~J. // Highly Redshifted Radio Lines,
                     ASP Conf. Series, 1999, V. 156, P. 121. \\
Levshakov~S.A., Varshalovich~D.A. // MNRAS, 1985, V. 212, P. 517 \\
Levshakov~S.A., Dessauges-Zavadsky~M., D'Odorico~S. // 2001, /astro-ph/0105529. \\
Petitjean~P., Srianand~R., Ledoux~C. // 2000, /astro-ph/0011437. \\
Pettini~M., Bowen~D. // 2001, /astro-ph/0104474. \\
Puy~D., Signore~M. // NewA, 1997, V. 2, P. 299. \\ 
Srianand~R., Petitjean~P., Ledoux~C. // 2000, /astro-ph/0012222. \\
Webb~J.K., Carswell~R.F., Lanzetta~K.M., Ferlet~R., Lemoine~M.,
        Vidal-Madjar~A., Bowen~D.V.) // Nature, 1997, V. 388, P. 250. \\

\newpage

\begin{figure}[h]
 \centering
  \includegraphics[height=190mm,bb=70 55 530 780,clip]{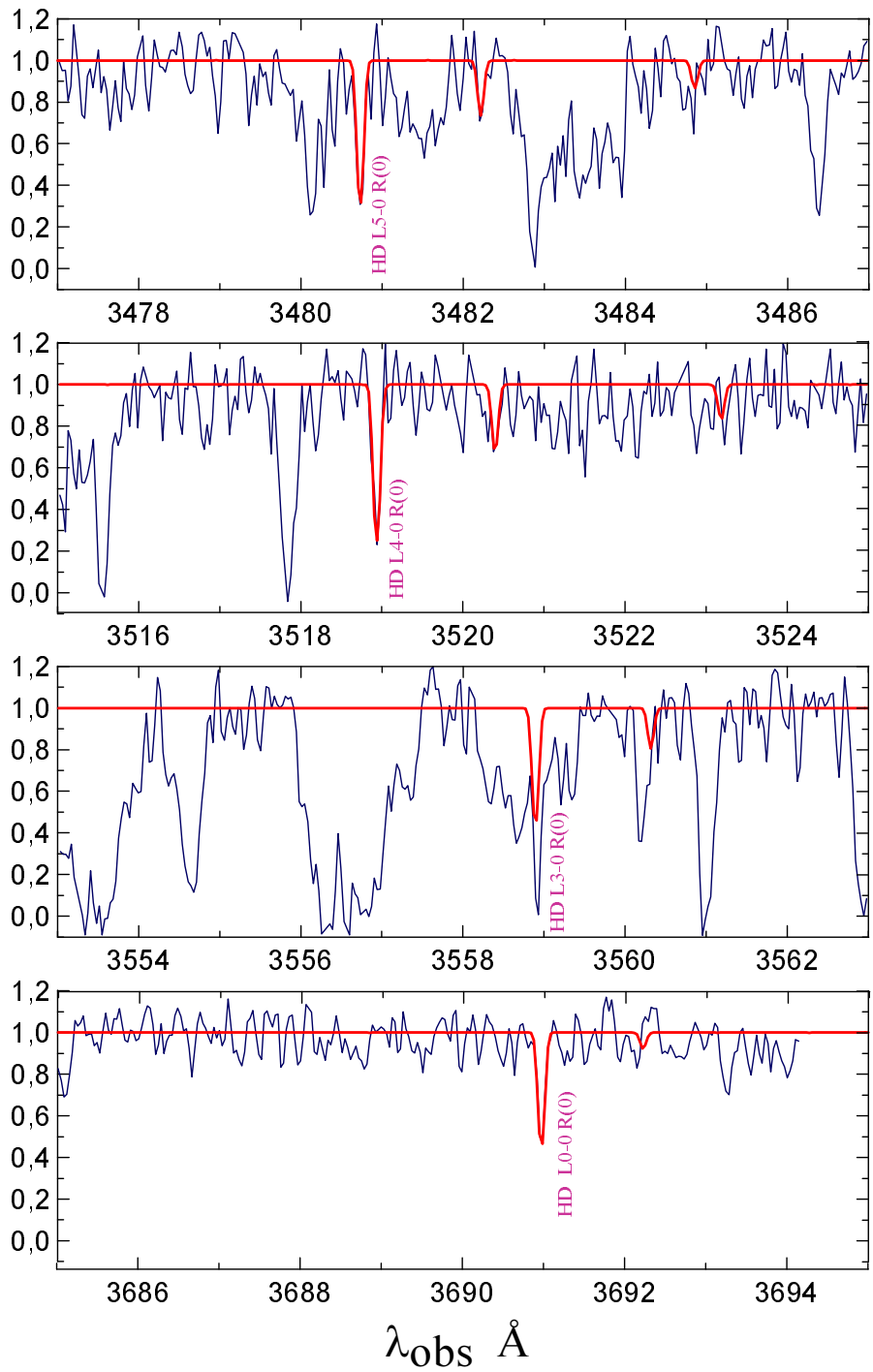}
\fontsize{12pt}{14pt}\selectfont
\caption{\Large ~~Fragments of PKS~1232+082 spectrum obtained 
by Petitjean~et~al. \protect\rule{15mm}{0mm}  (2000). 
The bold solid line shows the HD theoretical spectrum.} 

\end{figure}

\newpage

%----------- Table 1 --------------

  \begin{table}[t]
  \fontsize{12pt}{16pt}\selectfont
  \caption{\Large \protect\raggedright ~~~~~~Parameters of the identified lines of HD molecules
              \protect\rule{25mm}{0mm} in the spectrum of PKS~1232+082
	      in the system at $z_{abs}=2.3377$}
\vspace{7mm}

 \[
 \begin{array}{cclc}
 \hline
 \hline
 \noalign{\smallskip}
 \mbox{~~~~Transition~~~~} & \;\;\;\; \lambda_{lab}, \, \mbox{\AA} \;\;\;\; & \;\;\;\; \lambda_{obs}, \, \mbox{\AA} \;\;\;\; & \;\;\;\; z_{abs} \;\;\;\; \\
 \noalign{\smallskip}
 \hline
 \noalign{\smallskip}
 \mbox{L 5-0   R(0)} & 1042.854 & 3480.729(3) & 2.337695 \\
 \mbox{~~~~~~~~R(1)} & 1043.299 & 3482.219(9) & 2.337699 \\
 \noalign{\smallskip}
 \hline
 \noalign{\smallskip}
 \mbox{L 4-0   R(0)} & 1054.298 & 3518.935(4) & 2.337704 \\
 \mbox{~~~~~~~~R(1)} & 1054.734 & 3520.397(12) & 2.337711 \\
 \noalign{\smallskip}
 \hline
 \noalign{\smallskip}
 \mbox{L 3-0   R(0)} & 1066.279 & 3558.923(3) & 2.337703 \\
 \mbox{~~~~~~~~R(1)} & 1066.706 & 3560.34^b & ...  \\
 \noalign{\smallskip}
 \hline
 \noalign{\smallskip}
 \mbox{L 2-0   R(0)} & 1078.835 & 3600.83^b & ... \\
 \mbox{~~~~~~~~R(1)} & 1079.248 & 3602.21^b & ... \\
 \noalign{\smallskip}
 \hline
 \noalign{\smallskip}
 \mbox{L 1-0   R(0)} & 1092.006 & 3644.79^b & ... \\
 \mbox{~~~~~~~~R(1)} & 1092.404 & 3646.12^b & ... \\
 \noalign{\smallskip}
 \hline
 \noalign{\smallskip}
 \mbox{L 0-0   R(0)} & 1105.845 & 3690.972(2) & 2.337694 \\
 \mbox{~~~~~~~~R(1)} & 1106.221 & 3692.23     & ... \\
 \noalign{\smallskip}
 \hline
 \end{array}
 \]
 { $^b$  \mbox{HD line is blended} }
 \vspace{5cm}
 \end{table}

\end{document}